# A Novel Framework for DDoS Detection in Huge Scale Networks, Thanks to QoS Features

HamedRezaei[a,c], NimaGhazanfarimotlagh[a,d], YaghoubFarjami[b,e], Mohammad Hossein Yektaei[b,f]

[a]School of Engineering, Faculty of Information Technology, University of Qom-Iran. Alghadir Boulevard

[b]assistant Professor, School of Engineering, Faculty of Information Technology, University of Qom-Iran. Alghadir Boulevard

*Abstract--* **It is not been a long time since the advent of cloud-based technology. However, in this short period of time several advantages and disadvantages have been emerged. This is a problem solving technology with some threats as well. These threats and potential damages are not only limited to the cloud-based technologies, but they have always been against computer network infrastructures. One of these examples is Distributed Denial-of-Service (DDoS) intrusion which is of course one of the most complex and the most dangerous types of attacks. The impact of this type of attack, due to its powerful nature, is much higher on cloud systems since in case of occurrence, the service providers lose their services completely as well as their reputation and loyal customers. This, apparently, can even lead to the collapse of the stock and other destructive consequences. On the other hand, due to the properties of cloud service providers including large-scale infrastructures, DDoS intrusion detection algorithms need high sensitivity, innovation, and general improvements. Traditional structures of DDoS attack detection algorithms are designed for small-scale networks or at most for application camps. Lack of efficient algorithm is seemingly apparent for the large-scale networks. Therefore, in this context we utilize standard methods as well as a proposed hybrid protocol which is more appropriate in connection with cloud structures in order to detect DDoS attacks.**

*Keywords--* Cloud Computing, DDoS Attacks, Intrusion Detection Algorithms, Large-scale networks

## I. Literature Review

Many studies have been performed on cloud computing service providers. One of the most important sections of these studies is to review different challenges in these structures. For instance, a research has been conducted by the IDC in 2009, and after a full investigation among users, they have come to the conclusion that the most important challenge from the perspective of users of cloud services is security *[1]*. On the other hand, regarding the research performed by Lockheed Martin *[2]*, a major challenge after data security is Intrusion Detection in the cloud structures. So we can guess how critical and important communication security is even in this special case.

On the other hand, several studies have performed on the centralized threats of cloud structures. Cloud services and other network technologies may have many threats as well *[3]*. For example, we can mention the forgery attacks, unauthorized access attacks, and DDoS.

In this research we focus on DDoS attacks against cloud computing service providers. In these attacks, the attacker tries to fully utilize the resources required by users in order to prevent their access. These types of attacks can be occurred either in a distributed or a specific resource *[5]*. For example, in a cyber-attack occurred in 2011 against the Amazon EC2 service, Sony Online Entertainment segment was impaired and unavailable *[5]* which also caused in enormity losses.

With the emergence of cloud technology, researchers' attention in availability is increased much more rather than in information security *[6]* with regard to the fact that the basis of this technology is the high availability. Thus, several algorithms and methods are proposed to prevent attacks against availability *[3, 4, 6, 7, 8, 9, 10, 11, and 12]* each has proposed his own algorithm structure. For instance, *[13]* and *[14]* put ways of detecting an attack together and compared themgenerally. In addition, *[15]* evaluated the quality of the methods in intrusion detection alerts. Apparently, many techniques have been proposed for intrusion detection and they have all been studied by other researchers.

## II. Previous works

### 2.1 Artificial Neural Network (ANN) based IDS

The main purpose of artificial neural networks in DDoS detection of attacks, however, is proper classification of infected and healthy packets *[16]*. Many authors have presented their research in this field. For example, *[17]* have presented a more accurate IDS system with prediction and examination using more number of layers. Afterwards, [16] also presented a distributed method to achieve higher detection rate. However, the method provided by *[18]* is one of the few methods used in detection of DDoS attacks in cloud service providers. This method is a subcategory of disorder-based methods and requires more training time, but it can detect and neutralize most of the attacks.

### 2.2 Fuzzy Logic based IDS

These algorithms are designed to cope with any types of attacks, and are not specific to a particular type *[19]*. However, these technologies solely do not present a high level of impact and if they are combined with other methods of attacks detection, they would be more powerful.

For example, mixing with ANN methods can lead them to a great reduction in training time and thus increase the efficiency of the algorithm *[3]*.

*2.3 Association rule based IDS*

These types of intrusion detection algorithms can only detect attacks already exist in their database *[3]*, and if any undiagnosed attacks happen, they are unable to detect. One of the best methods using these algorithms is presented by *[20]*. This method along with special rules used to specify attacks takes benefits of data mining techniques in order to update rules. Since its time consuming nature is the main problem *[3]*, continuous researches are investigating on this algorithm, but due to the long lifetime of this method, any progress is not seemed to be significantly achieved.

*2.4 Genetic Algorithm (GA) based IDS*

These methods are not able to detect attacks by themselves; therefore, they should be combined with other methods to enhance their performance [21]. Many researches have been carried out in combination of these algorithms with other methods *[21]* and *[22]*. These algorithms can be perfectly used in cloud environments due to their nature of improving potential.

### III. PROPOSED ALGORITHM

The algorithm presented in this paper uses the logic of distributed intrusion detection systems. This means that several sections examine the probability of intrusion detection and in case they ensure the DDoS attack is happening, they prevent the damaging effects in cooperation with each other. In the applied algorithm, two important points are addressed; first, almost the same as other intrusion detection algorithms, it uses a training phase, a real world phase, or a test. Second, the scale of the intrusion detection in this algorithm is the bandwidth usage *[25]*. Of course there are other scales, but the most appropriate scale to detect DDoS attacks is the bandwidth that is used. Our considered algorithm consists of several sections which are connected to each other in a completely distributed way. These components are shown in the figure below:

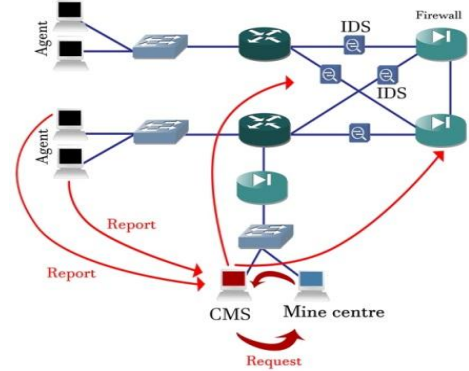

**Figure 1. General scheme of the algorithm**

- A small software agent which will be installed on each virtual server
- Control center in each zone of the cloud environment that has a full control over all components
- Intrusion pattern detection center
- Firewall
- IDPS

Each of these five sections has its own responsibility coordinating with control center. Following is the brief introduction to the responsibilities of each section.

*3.1 Mining Center*

The purpose of this center is to investigate about the unusual events that are not previously detected as an attack. In such circumstances, the control center asks this unit to examine packets. If a new pattern of intrusion is observed, it is stated. The algorithm applied in this section is presented by *[24]* which can automatically create attack rules based on the inbound traffic. This algorithm is one of the most intelligent algorithms of intrusion detection which is able to easily examine the packets and detect the potential or even actual intrusions to prevent DDoS attacks.

*3.2 CMS (Central Management Service)*

This is the algorithm's mastermind and coordinator of all activities. This center is responsible for:

1) Receive a full report of high consumption users from the software agent
2) Send the traffic to the mining center in order to detect intrusion pattern
3) In case of a new attack detection, send the new rule to the IDPS
4) Receive increasing bandwidth usage alert from the software agents
5) Receive the second-level alert (described in the next section) from the software agent
6) Deliver necessary instructions (e.g. police) to the firewall
7) In case of changing the bandwidth of a virtual machine on the firewall, report the value to the same machine
8) Receive the third-level alert from the control center

*3.3 Software Agent*

This agent is responsible for several tasks, including the real-time report of traffic load on the virtual servers whereby different alerts will be sent to the control center. Another important responsibility of this agent is to identify the most high-consumption users in the system. This happens because in case of attacks, we can distinguish our important users from other users.

It's of high importance state that this agent only responds to the control center and does not communicate with other sections of the protocol. The next important fact is the main cause of putting this agent on virtual machines, instead of the main server which can be considered as a host for several virtual servers, is that usually the intrusions are occurred against a specific server with a specific IP address. As another fact, the bandwidth used by the virtual machines is way less than the usage of the main server. So it makes sense that this intelligent software agent is installed separately on each VM and only control the same VM.

*3.4 IDPS*

The role of IDPS in DDoS intrusions prevention algorithms is definitely not negligible. One of the best examples of these tools is "Snort" which unfortunately does not have sufficient capability to detect complex and new attacks. Therefore, equipping this service with the Artificial Intelligence (AI) structures, we can significantly enhance our ability to detect intrusions. The first important fact about this service is that it also communicates only with the main control center, and the second is that it is placed exactly behind main firewalls.

*3.5 Firewall*

In the proposed algorithm, firewall is responsible for the following roles in addition to its primary role as an unauthorized packet deterrent:

- Receiving Quality of Service (QoS) commands from the control center
- Policing the traffic

IV. HOW DOES THE ALGORITHM WORK

All the five mentioned sections should be fully functional and dynamic and also should communicate with the control center. This structure is in a form of header between the IP layer and the transport layer, and will be placed as follows. In fact, it is the same as the IPsec header on the 3rd layer and carries its own information. Data section also includes information from packets which are sent from the agents to the control center. The list of high-consumption users is placed in this section. 3 bits is reserved for source section representing the packet generator, i.e. 000 indicates control center, 001 indicates firewall and so forth. Device ID needs 5 bits to clarify the packets, like Alloha, Alert levels1&2&3, ACK, and so on. Sequence Number shows the sequence in which packets are sent. Authentication Key is the key between sender machine and control center. Next Protocol Number represents the number of protocol in layer 4, i.e. tcp-7. It should be mentioned that one of the most important benefits of this algorithm is the lack of progress in the higher levels and all communications and messages are taking place in the same sector. Packet header is shown in the figure below:

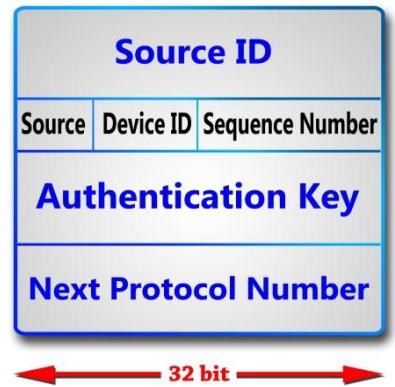

**Figure 2: Packet Header**

Reminding the fact that the inbound traffic to a network card is principally following the Normaldistribution function *[26]*, led us to produce this algorithm.

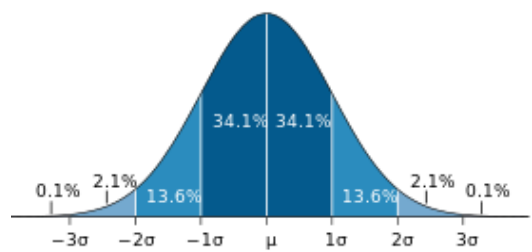

**Figure 3: General scheme of the Normal Distribution Function**

Most noticeable data will be in the interval of $\mu \pm 2\sigma$. We use the Aging Algorithm to predict the amount of bandwidth. The general form of the algorithm is as follows:

$$S_n = xS_{n-1} + (1-x)T_{n-1} \quad ,(1)$$

This algorithm is one of the most important and oldest methods of predicting the lifespan of processes in operating systems [23] and we can find the amount of approximate bandwidth usage for current period of time using (1). Thus we have:

$$\alpha = S_n \quad ,(2)$$

According to the predictions, we expect not to see the bandwidth usage out of the considered range in a normal usage, and to formulate this issue, we have utilized 3 levels of alerts. In this formula, $\alpha$ is the predicted amount, and $\beta$ is calculated as follows. Since we need 3 levels of alerts, it's apparent that we need $\beta$ to be:

$$\beta < \frac{100 - \alpha}{3} \quad ,(3)$$

Therefore, the boundary point for $\beta$ is:

$$\beta = \frac{100 - \alpha}{3} \quad ,(4)$$

There is another consideration; in the case that $\beta$ is greater than real deviation of consumption ($\sigma$), it implies that we do not need the extra part of $\beta$ and we have:

$$\beta = \sigma \quad ,(5)$$

*4.1 First Level Alert*

If $RealLoad \geq \alpha + \beta$, software agents installed on the virtual machines start storing the incoming packets into a database. This is performed because in the case that we decide to send packets to the mining center, good samples are already provided in our database.

*4.2 Second Level Alert*

If $RealLoad \geq \alpha + 2\beta$, then network traffic is passed over a significant point. To acknowledge the existence of a second level alert, the storage file is sent to the control center through the software which includes incoming packets after reaching the first level alert. This is performed by the agent itself. The data is sent to the control center. Afterwards it is examined in mine center in order to detect possible attacks.

*4.3 Third Level Alert*

If $RealLoad \geq \alpha + 3\beta$, a special exceeded case occurs requiring urgent action to prevent excessive bandwidth usage. Since the intrusion detection is depended on the running algorithm in data mining center, we propose a special solution for this situation.

It is expected that the mining center extract the pattern of probable attack having packets received by the second and third level alert. In the following section, two possible situations are discussed.

## V. DIFFERENT ATTACK SCENARIOS

*5.1 First Scenario: The pattern of attack is detected in a proper time by mining center*

The first probable scenario of attack detection is by detecting heterogeneity in the exchanged traffic and sending it to the data mining center, the intrusion pattern is obtained clearly within a defined period of time. Data mining center sends this pattern formally to the control center after detection. Control center also converts this pattern to an understandable form for the Snort, therefore, the pattern is recorded along with other patterns of attacks. Hereinafter, this particular type of intrusion to VMs will be stopped afterwards. The value of $\alpha$ is set to its previous status before attack happening in order not to lose prediction accuracy.

*5.2 Second Scenario: The pattern of attack is NOT detected in a proper time by mining center*

This scenario as the previous one has the same general algorithm. The difference is about the data mining center not being able to detect the pattern of attack within the specified period of time, and needs more time to evaluate the traffic. This has two reasons; first, it is likely that the attack is not happened at all and the increase of bandwidth usage is resulting from the use of legitimate users. Second, it is likely that the attack is happened, but detection of its pattern is not possible in less than 30 seconds [24]. In this case, one of the best possible solutions is Policing. The important question is: when the operation starts? Certainly, users are allowed to freely use the service unless they have not entered the third level of alerts, but upon entering the third warning, all users have to be policed to the extent that the total bandwidth usage is placed on the predicted $\alpha$.

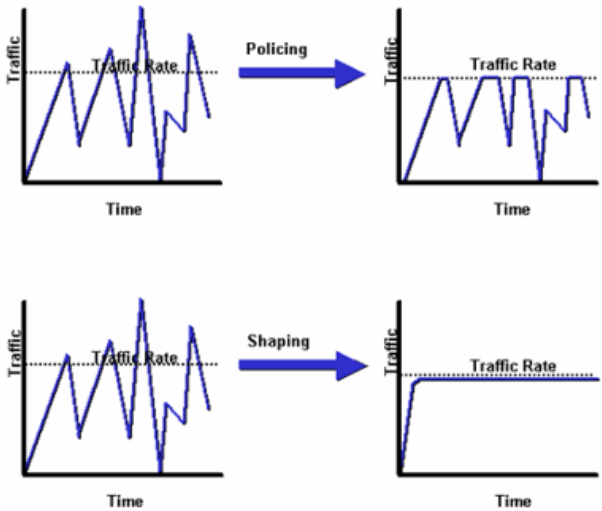

**Figure 4: Policing & Shaping**

It is expected that during a period of up to 2 minutes *[23]*, the attack is detected. In this period oftime, the Policing percentage is reduced every 30 seconds and finally the total bandwidth is released. This process will be checked continuously

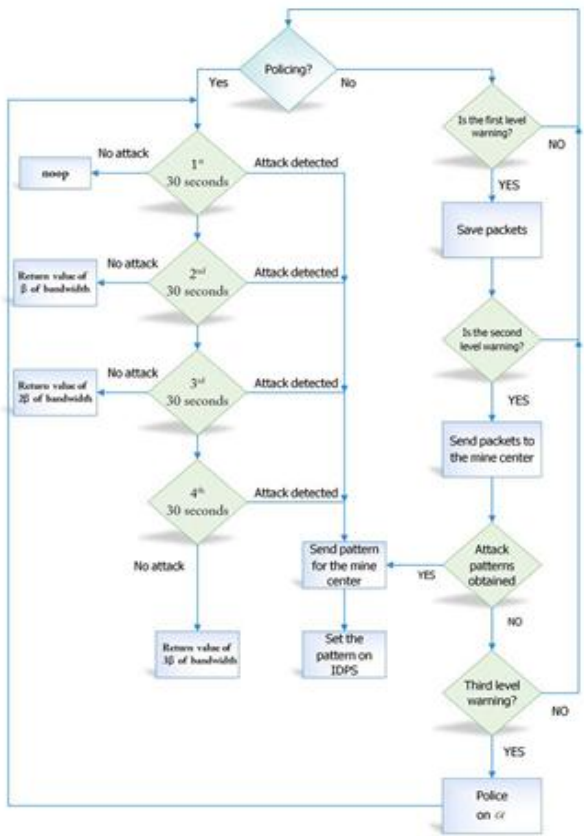

**Figure 5: Running Applications Flow Chart in the Perspective of control**

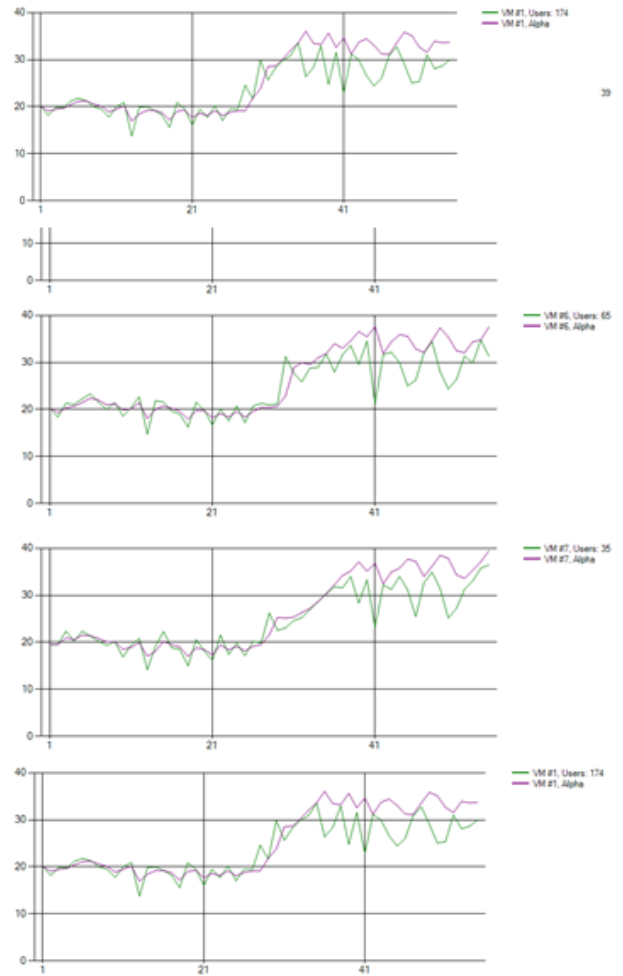

**Figure6. Bandwidth Usage Rate against Alpha, on one of the VMs**

Finally, intable 2, the algorithm presented in this paper is compared with other well-known algorithms in this field.

## VI. PROPOSED ALGORITHM OUTPUTS

Testing the proposed algorithm, we have designed a simulator application in which various outputs are illustrated in this section:

### 6.1 Alpha Output (α) and Data Usage

For all diagrams, the horizontal axis represents the time intervalfrom the beginning of the algorithm. After a while, a probable attack is performed against the system.Outputs show that the system tolerates a simulated DDoS attack performing desired actions.

Here, the vertical axis represents total percentage of bandwidth usage on different VMs.

### 6.2 Police Rate

The vertical axis represents total percentage of police per total bandwidth usage.

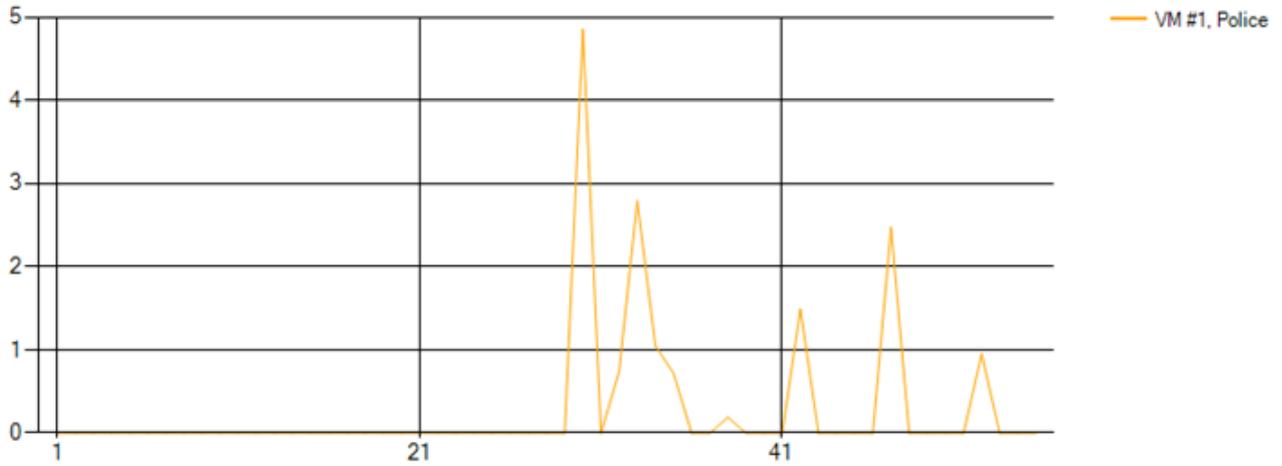

Figure7. Police Rate on one of the VMs

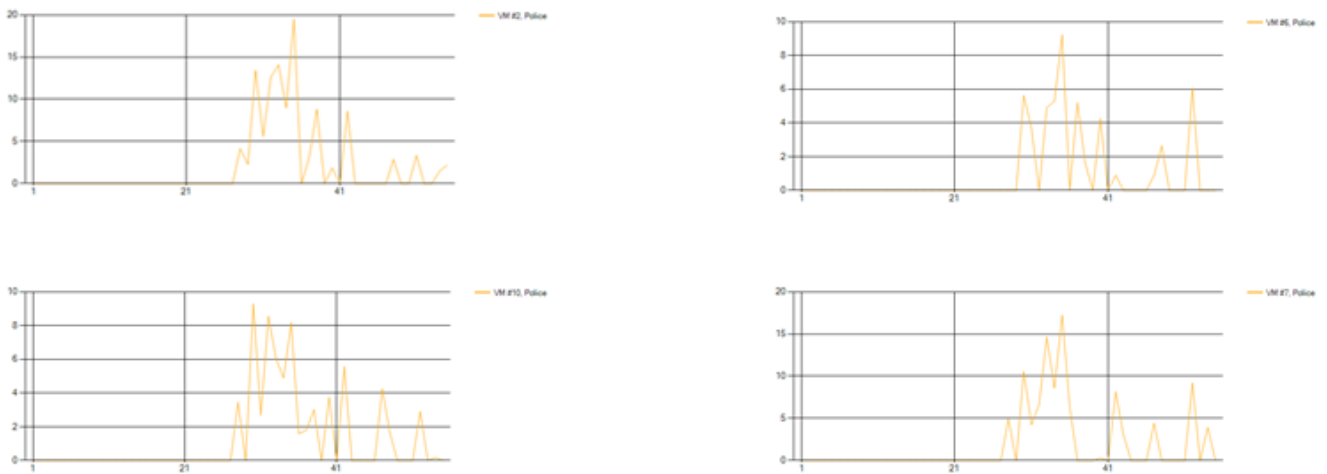

### 6.3 Statistical of Attackers

The vertical axis represents the percentage of attackers among total number of users.

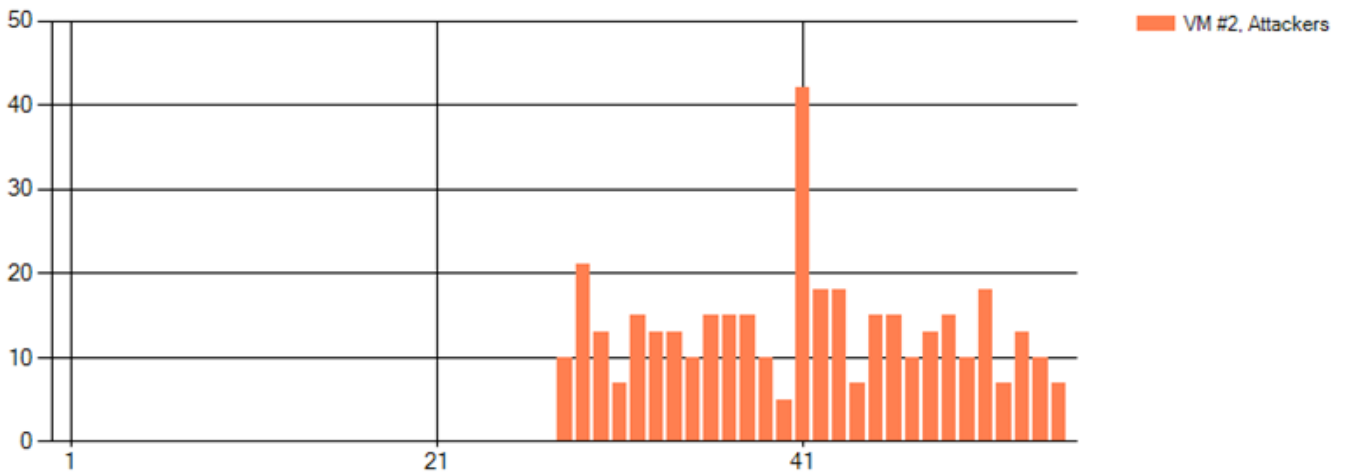

Figure8. Attacks Rate on one of the VMs

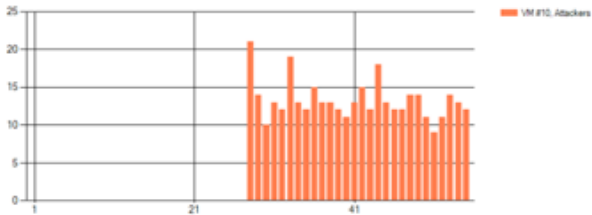
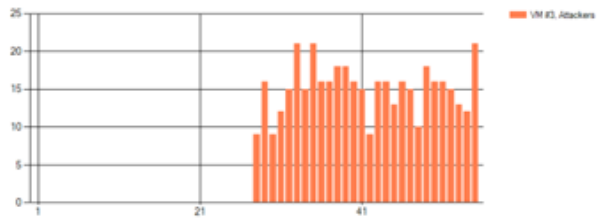
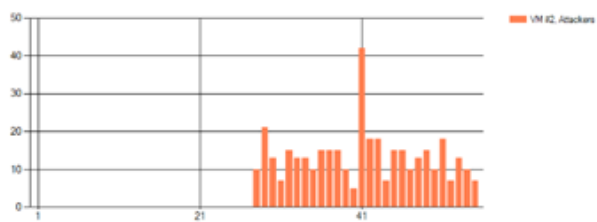
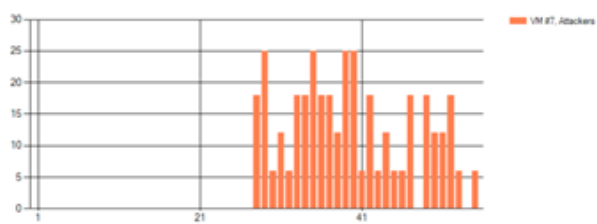

Apparently, the behavior of the algorithm for different machines and different traffics is quite appropriate; furthermore, in addition to allowing users to use their maximum possible bandwidth, it also avoids any violation of the rules.

*6.4 Bandwidth Preservation for high consumption users:*

There are different labeling protocols, namely CoS, DSCP, IPP. Being more recent and requirement covering, we selected DSCP for our algorithm. The following table shows general properties of this labeling protocol:

**Table1.**
**Different properties of DSCP**

| Queue Class | Low Drop Probability | Medium Drop Probability | High Drop Probability |
|---|---|---|---|
| | Name/ Decimal/ Binary | Name/ Decimal/ Binary | Name/ Decimal/ Binary |
| 1 | AF11/ 10/ 001010 | AF12/ 12/ 001100 | AF13/ 14/ 001110 |
| 2 | AF21/18/ 010010 | AF22/ 20/ 010100 | AF23/ 22/ 010110 |
| 4 | AF31/ 26/ 011010 | AF32/ 28/ 011100 | AF33/ 30/ 011110 |
| 5 | AF41/34/ 100010 | AF42/ 36/ 100100 | AF43/38/ 100110 |

In our proposed algorithm, the control center is to put high consumption users in a priority of 111000, so that these types of packets are prohibited after all other packets are. With this in mind, those users who have already been registered in the accounting service are allowed to continue using service freely with the priority of AF41. Other users are circumstantially policed, according to network administrator or other principal's decision.

*6.5 Comparison of Proposed Algorithm with Other Algorithms*

Any efficient algorithms require 8 following conditions:[14]

*R1:* Does it manage independent multi-layered environments and large-scale and dynamic data processing?

Cloud computing environments are different from other network infrastructures. Because in cloud computing, which involves large scale networks and systems, it is crucial to maintain some changes automatically without human intervention.[14]

*R2:* Does it detect all different types of intrusions with the minimum rate of positive false alarm?

It detects variety of attacks with least False Positive Rates Due to the growth of attacks, complexity and unpredictability, it is necessary for the system to recognize new attacks and their vulnerable intention to choose the best response according to the risk severity and proper prevention strategy*[14]*.Therefore, a good detection method must detect all types of attacks without consuming huge amount of resources.

*R3:* Very fast detection and prevention

Sharp detection and prevention is a very important enabling factor for CIDPS since it affects the whole system performance and is crucial to deliver the pre-agreed QoS.*[14]*

*R4:* Automatic self-adaptive

One of the most important factors of a successful detection method is being self-adaptive and responding to the environment and various types of attacks in a positive way. An IDPS should configure itself and be adaptive to configuration changes as computing nodes are dynamically added and removed. *[14]*

*R5:* IDPS scalability

A CIDPS should be scalable in order to efficiently handle the massive number of network nodes available in cloud and their communication and computational load. *[14]*

*R6:* Certainty

A CIDPS should be able to provide and maintain an acceptable level of service in the occurrence of faults, be highly reliable and deliver very high uptime services imposing minimal overhead.

A CIDPS should not only ensure real time performance, but also ensure that the deterministic nature of network is not adversely affected. *[14]*

*R7:* Independent IDPS synchronization

The IDPS system must be completely unified to respond all the attacks in an efficient way. While each subsystem operates and detects intrusions and anomalies independently, their information and activities must be synchronized in order to recognize distributed and concurrent attacks, apply appropriate response or modify a particular component subsystem or the whole network configuration, and adopt proper prevention strategy.*[14]*

*R8:* Resistance to attacks

An IDPS must protect itself from unauthorized access or attacks. A CIDPS should get benefits of authenticating network devices and IDPSs mutually, authenticating the administrator and auditing his actions, protecting its data, and blocking any loopholes which may create additional vulnerabilities. *[14]*

In figure 10 a brief analogy between our proposed algorithm and other successful algorithms with regard to these eight yardsticks is provided.

**Table2.**
**Algorithm comparison**

| Comparison of some detection algorithms [14] | | | | | | | | | |
|---|---|---|---|---|---|---|---|---|---|
| reference | R1 | R2 | R3 | R4 | R5 | R6 | R7 | R8 |
| Proposed algorithm | √ | N/A | √ | √ | √ | √ | √ | √ |
| [27] | × | √ | × | √ | √ | √ | × | √ |
| [28] | × | N/A | N/A | √ | √ | √ | × | N/A |
| [29] | × | p | × | √ | √ | × | N/A | √ |
| [30] | × | p | × | N/A | N/A | × | × | √ |
| [31] | × | √ | p | N/A | N/A | p | N/A | p |
| [32] | × | √ | × | N/A | N/A | √ | N/A | × |
| [33] | p | √ | × | √ | √ | √ | × | √ |
| [34] | × | √ | N/A | √ | √ | × | N/A | √ |
| [35] | √ | √ | √ | √ | N/A | p | × | √ |
| [36] | √ | p | × | √ | √ | p | N/A | N/A |

P= partially   X=doesn't support   √= supports   N/A= not applicable

VII. CONCLUSIONS AND RECOMMENDATIONS

After all investigations, we come to the fact that it is possible to increase the power of intrusion detection as well as using the desired detection algorithm and prevent a network crash. Therefore, referring to the mathematical aspects of the proposed algorithm, it is feasible to fully utilize it asa supplemental defense besides other intrusion detection algorithms. Since we have attempted to achieve our objective to obtain a safe and on-line protocol, this protocol can be applied to not only small or medium-case networks, but also very large-scale networks as well. The reason is that its centralized mechanism will allow the entire network to be connected and synchronized with a singlecontrol center and their data is periodically checked against any intrusions.As future works, there can be many researches to be conducted; from the methods of calculating predicted results to the means of protocol users talk to the multi-cast members, theauthentication key between virtual machines and the control center, and a more complicated method to use more CMSs avoiding single point of failure (SPOF).